\documentclass[preprint]{revtex4}
\usepackage{graphicx}
\usepackage{subfigure}
\usepackage{caption}
\usepackage{amsmath}
\usepackage{array}

\newcommand{\beq}{\begin{equation}}
\newcommand{\eeq}{\end{equation}}
\newcommand{\beqa}{\begin{eqnarray}} 
\newcommand{\eeqa}{\end{eqnarray}}
\newcommand{\blgn}{\begin{aligned}}
\newcommand{\elgn}{\end{aligned}}

\newcommand{\la}{\langle}
\newcommand{\ra}{\rangle}
\newcommand{\da}{\dagger}
\newcommand{\de}{\delta}
\newcommand{\f}{\frac}

\newcommand{\om}{\omega}

\newcommand{\pa}{\partial}

\newcommand{\td}{\tilde} 
\newcommand{\nn}{\nonumber \\}

\newcommand{\eq}[1]{Eq. \eqref{#1}}

\begin{document}

\title{Non-local Geometry inside  Lifshitz Horizon}

\author{Qi Hu$^{1,3}$ and Sung-Sik Lee$^{1,2}$\\ 
\vspace{0.3cm}
{\normalsize{$^{1}$Perimeter Institute for Theoretical 
Physics, Waterloo, ON N2L 2Y5, Canada}}\\
{\normalsize{$^{2}$Department of Physics $\&$ Astronomy, 
McMaster University, Hamilton, ON L8S 4M1,Canada}}\\
{\normalsize{$^{3}$Department of Physics and Astronomy, University of Waterloo, Waterloo, ON N2L 3G1, Canada}}\\
}

\date{\today}

\begin{abstract}
Based on the quantum renormalization group,
we derive the bulk geometry
that emerges in the holographic dual 
of the fermionic $U(N)$ vector model
at a nonzero charge density.
The obstruction that prohibits the metallic state
from being smoothly deformable to the direct product state 
under the renormalization group flow
gives rise to a  horizon
at a finite radial coordinate in the bulk.
The region outside the horizon 
is described by the Lifshitz geometry
with a higher-spin hair 
determined by microscopic details of the boundary theory.
On the other hand, the interior of the horizon
is not described by any Riemannian manifold,
as it exhibits an algebraic non-locality.
The non-local structure inside the horizon 
carries the information on
the shape of the filled Fermi sea.
\end{abstract}

\maketitle

\section{Introduction}

Black hole horizon
hosts tensions among  
basic principles of physics
established within the framework of local field theories,
and understanding what is behind the horizon  
may hold the key to 
the resolution\cite{Hawking1975,THOOFT1990138,PhysRevD.48.3743,0264-9381-26-22-224001,Almheiri2013}.
Given that there is so far no non-perturbative definition of quantum gravity
except through the AdS/CFT correspondence\cite{Maldacena:1997re,Witten:1998qj,Gubser:1998bc},
it is desired to have an access to the interior of horizon
via boundary field theories\cite{
PhysRevD.67.124022,
PhysRevD.75.106001,
Papadodimas2013,Kabat2014,PhysRevLett.112.051301}.
However, probing the interior of horizon
may require a full microscopic theory of 
the bulk beyond the local field theory approximation
whose validity can not be taken for granted near horizon.

Quantum renormalization group (RG)
provides a microscopic prescription to derive holographic duals for general field theories\cite{Lee2012,Lee:2012xba,Lee:2013dln}.
In quantum RG, 
renormalization group flow is mapped to a dynamical system,
where the action principle replaces classical beta functions.
The sources for a subset of operators (called single-trace operators)
become dynamical variables
whose fluctuations encode the information about all other operators
which are not explicitly included in the RG flow.
The resulting bulk theory generally includes dynamical gravity\cite{Lee:2012xba,Lee:2013dln}
and gauge theory\cite{doi:10.1142/S0217751X13501662,Bednik}
because the background metric and gauge field 
that source the energy-momentum tensor and a conserved current
are promoted to dynamical variables.
From the bulk perspective,
this accounts for the fact that 
multi-trace operators are generated 
once bulk fields for single-trace operators are integrated out\cite{Heemskerk:2010hk,2011JHEP...08..051F,2013JHEP...01..030K}.
Because the complete set of single-trace operators
include operators of all sizes in general, 
the bulk theory is kinematically non-local\cite{2015PhRvL.114c1104M,Lee:2012xba}.
In the presence of non-local dynamical degrees of freedom,
locality in the bulk is a feature 
that is determined dynamically  
rather than a kinematical structure put in by hand.
This provides a natural setting
to study horizon within
the framework of RG flow.

In quantum RG, 
horizon corresponds to 
a Hagedorn-like dynamical phase transition
where non-local operators proliferate
at a critical RG scale\cite{Lee2016}.
This is best understood in terms of 
quantum states defined on spacetime,
where RG flow of an action
is viewed as a quantum evolution
of the corresponding state 
generated by a coarse graining operator.
The coarse graining generator
projects the state associated with an action 
toward a reference state that represents an IR fixed point.
Whether the true IR physics of the theory is described 
by the putative IR fixed point or not 
is determined by whether the state can be smoothly projected to the reference state or not.  
Although a local action is mapped to a short-range entangled state,
the range of entanglement increases 
under the RG flow
as non-local operators are generated.
If the theory is in a gapless phase,
the quantum state can not be smoothly projected to a reference state
which represents a gapped state.
The obstruction manifests itself 
as a proliferation of non-local operators
at a critical RG scale.
This marks as a dynamical phase transition
whose order parameter is locality (or loss of locality),
and the critical point gives rise to a horizon in the bulk.

In an earlier work\cite{Lee2016},
the correspondence between critical phenomenon and horizon 
has been demonstrated in a boson model.
However, it is not possible to cross the horizon 
in the boson model because the critical point 
arises at an infinite RG scale. 
In this paper, we study a fermionic model with a nonzero charge density
which undergoes a phase transition at a finite RG scale 
associated with the chemical potential.
Since a horizon arises at a finite radial coordinate, 
one can go through it
to reach the interior via  RG flow.
While the outside of horizon is described by a Lifshitz geometry\cite{
PhysRevD.78.106005,
Balasubramanian2010,2010JHEP...12..002D,
2012arXiv1202.6635H},
the interior of the horizon exhibits a non-local geometry 
which can not be described by a Riemannian manifold.
We  show that the non-local structure inside the horizon 
is sensitive not only to
the universal long-distance properties of the boundary theory
but also to  microscopic details.

\section{Holographic dual for the fermionic vector model}

\begin{figure}[h]
\centering
	\subfigure[]{\includegraphics[width=1.8in]{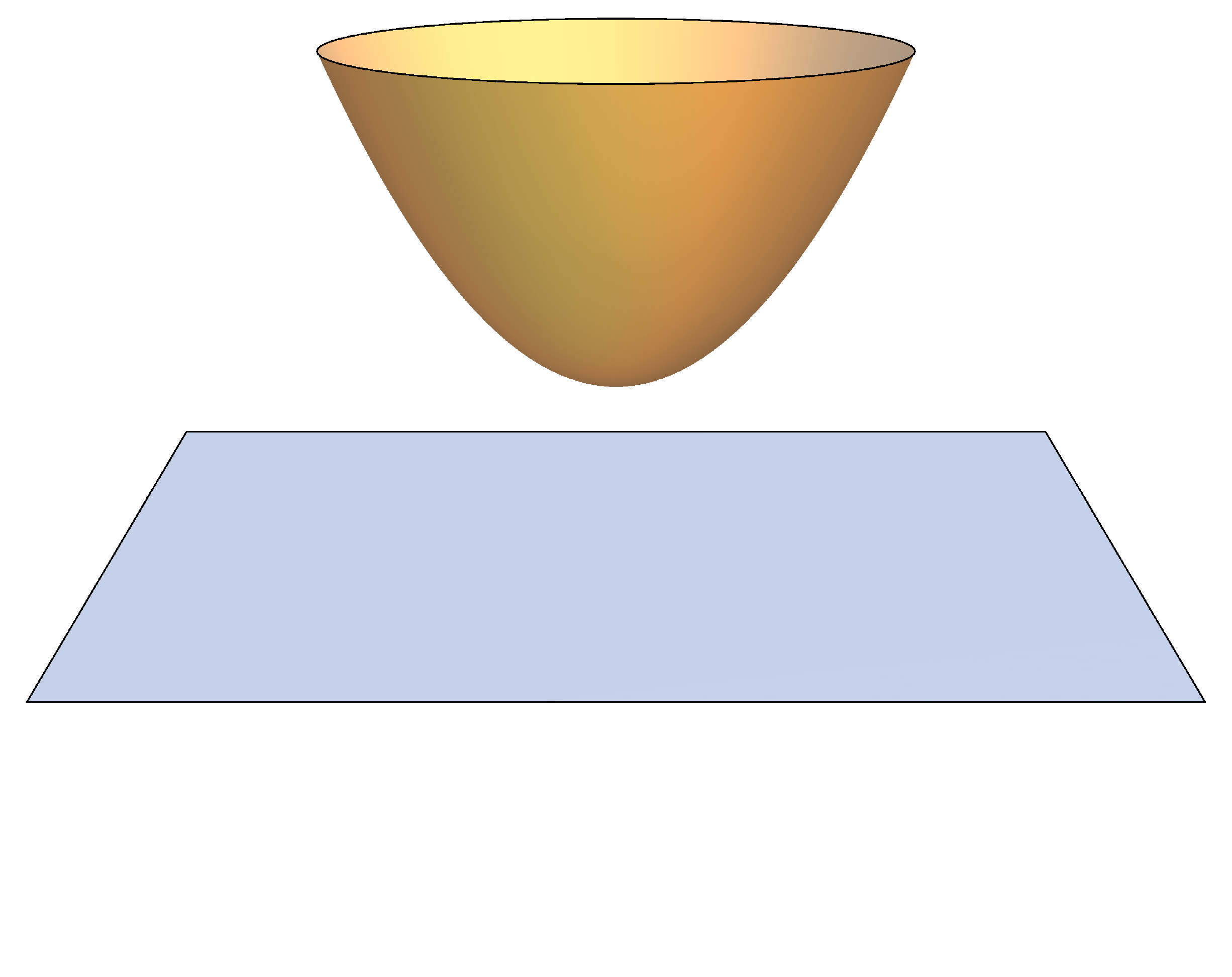}}
	\subfigure[]{\includegraphics[width=1.8in]{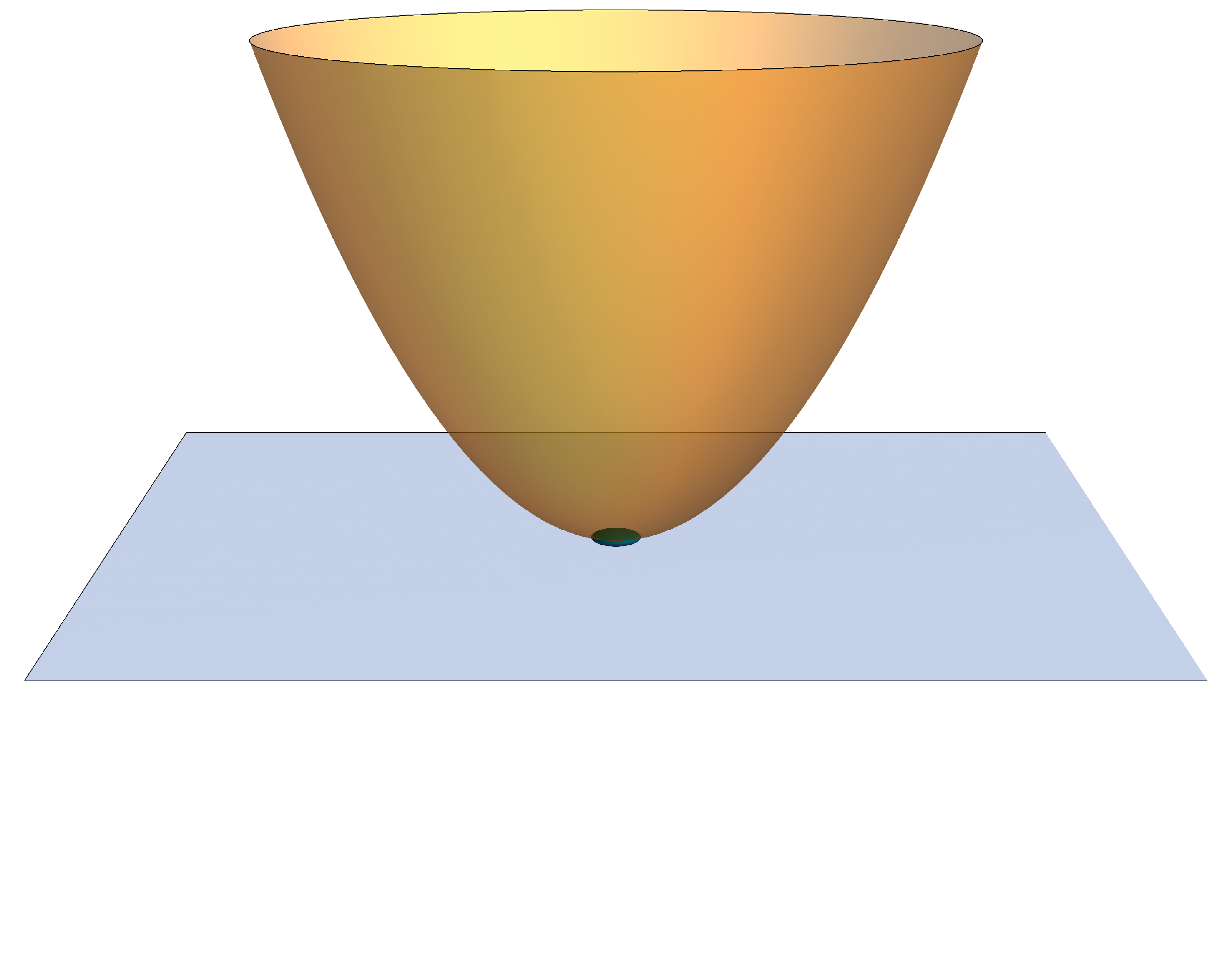}}
	\subfigure[]{\includegraphics[width=1.8in]{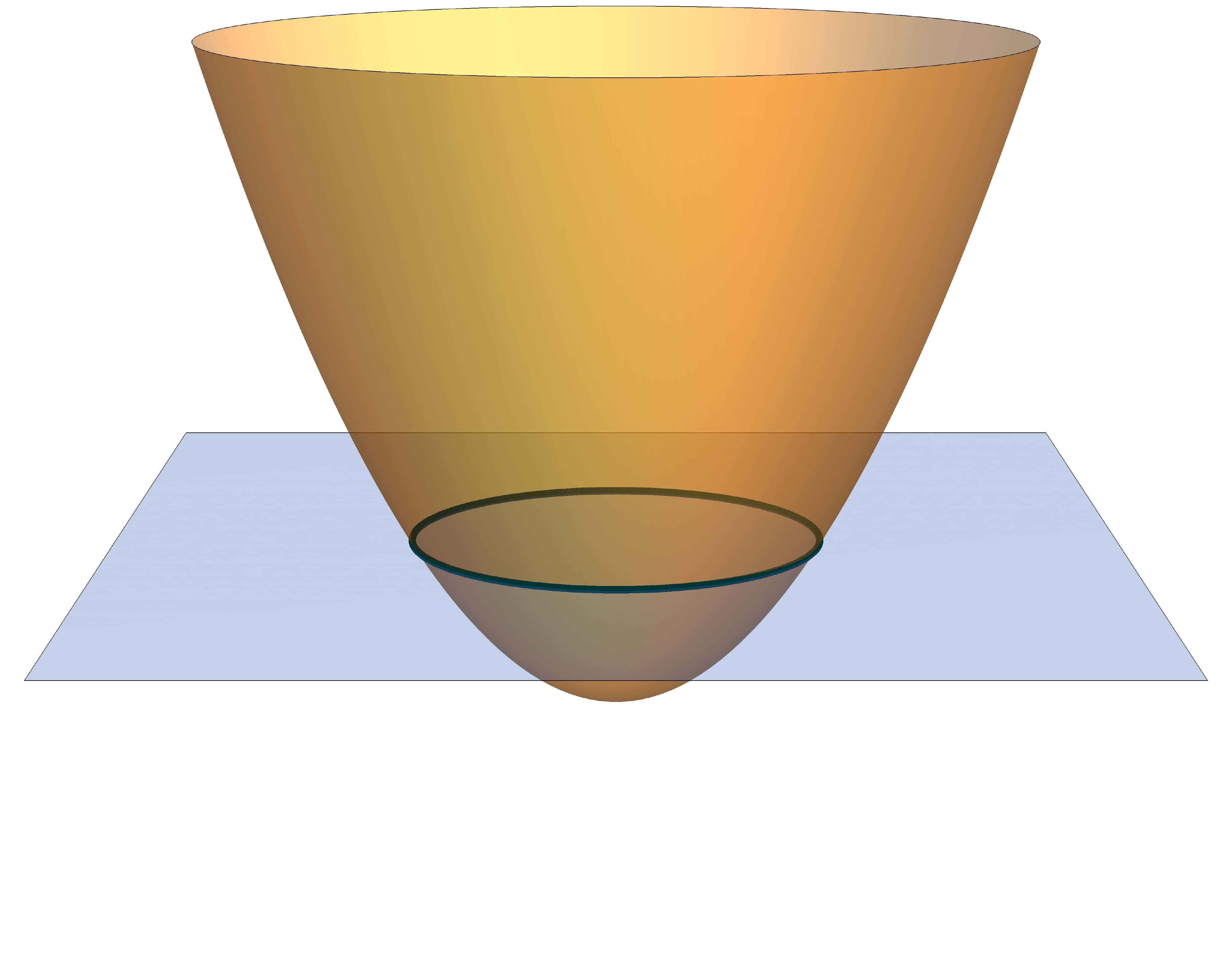}}
\caption{
The energy dispersion 
for (a) insulator, (b)  critical point and (c) metal
plotted in the two-dimensional momentum space.
The dot in (b) represents the bottom of the band
which touches the chemical potential at the critical point,
and the circle in (c) represents the Fermi surface.
}
\label{spectrum}
\end{figure}

We consider a $D$-dimensional fermionic U(N) vector model,
\begin{equation}
\mathcal S =\int d\tau d^{D-1} x~ 
\left[ \bar {\bm\psi}^a \partial_\tau {\bm\psi}^a 
+\nabla  \bar{\bm\psi}^a \cdot \nabla{\bm\psi}^a
- \mu  ~\bar {\bm\psi}^a  {\bm\psi}^a +\frac \lambda N ~(\bar {\bm\psi}^a \cdot {\bm\psi}^a)^2 \right].
\label{eq:ca}
\end{equation}
Here $\tau$ is the imaginary time,
$\bar {\bm\psi}^a$ and ${\bm\psi}^a$ are Grassmann fields with flavour $a=1,2,..,N$.
$\mu$ is the chemical potential for the $U(1)$ charge, 
and $\lambda$ is a quartic coupling. 
We regularize the field theory on a $D$-dimensional lattice as
\begin{equation}
\mathcal S=-\sum_{ij} t^{(0)}_{ij} (\bar {\bm\psi}_i \cdot {\bm\psi}_j)+ m \sum_i (\bar {\bm\psi}_i \cdot {\bm\psi}_i)+\frac \lambda N \sum_i (\bar {\bm\psi}_i \cdot {\bm\psi}_i)^2.
\end{equation}
Here $i,j$ indicate sites on the $D$-dimensional spacetime lattice.
$\bar {\bm\psi}_i \cdot {\bm\psi}_j \equiv \sum_a \bar {\bm\psi}^a_i  {\bm\psi}^a_j$
represents the set of bi-local single-trace operators, 
and  $t^{(0)}_{ij}$'s are hopping amplitudes on the lattice.  
The local chemical potential can be identified as
$\mu_i = \sum_j t^{(0)}_{ij} - m $,
where $m>0$ is assumed.
If the hopping is small compared to $m$,
the chemical potential is below the bottom of the band,
and the system is in the insulating state.
On the other hand, a finite charge density is generated
and the system becomes a metal
when the hopping is large.
The system goes through the insulator to metal transition
as the magnitude of $t_{ij}^{(0)}$ is increased, 
as is illustrated in Fig. \ref{spectrum}.
Our goal is to derive the bulk geometry
that emerges in each state.   
For other holographic approaches to related vector models, 
see Refs. \cite{2002PhLB..550..213K,
Das:2003vw,
Koch:2010cy,
Douglas:2010rc,
2013arXiv1303.6641P,
Leigh:2014tza,
2015PhRvD..91b6002L,
2014arXiv1411.3151M,
Vasiliev:1995dn,
Vasiliev:1999ba,
Giombi:2009wh,
Vasiliev:2003ev,
Maldacena:2011jn,
Maldacena:2012sf,
2014PhRvD..90h5003S}.

To set up the RG procedure,
we divide $\mathcal S$ into 
a reference action and a deformation.
The reference action is chosen to be the ultra-local theory
which describes the insulating fixed point,
\begin{equation}
\mathcal S_0= m \sum_i (\bar {\bm\psi}_i \cdot {\bm\psi}_i),
\end{equation}
and the rest is treated as a deformation,
\begin{equation}
\mathcal S_1=-\sum_{ij} t^{(0)}_{ij} (\bar {\bm\psi}_i \cdot {\bm\psi}_j)+\frac \lambda N \sum_i (\bar {\bm\psi}_i \cdot {\bm\psi}_i)^2.
\end{equation}
Following \cite{Lee2016}, 
we define quantum states associated with $S_0$ and $S_1$ as
\beqa
|S_0 \rangle & = &  \int D\bar\psi D\psi ~e^{-S_0[\bar\psi,\psi]} |\bar\psi,\psi \rangle, \nn
|S_1 \rangle & = &  \int D\bar\psi D\psi ~e^{-S_1[\bar\psi,\psi]} |\bar\psi,\psi \rangle,
\eeqa
where $|\bar\psi,\psi \rangle$ are the basis states labeled by the Grassmannian fields,
\beq
|\bar\psi,\psi\ra= \prod_{i,a} (1-\bar \psi_i^a  c_{1i}^{a \da})(1-\psi_i^a  c_{2i}^{a \da})|0\ra.
\eeq
The basis states are constructed from a Fock space
which can accommodate up to $2N$ auxiliary fermions at each site
on the $D$-dimensional spacetime lattice :
$c_{1 i }^{a \dagger}$
($c_{2 i }^{a \dagger}$)
 with $a=1,2,..,N$
 represents the creation operator
 of the fermions associated with $\bar \psi_i^a$ ($\psi_i^a$), 
 and $|0\ra$ is the vacuum annihilated by $c_{\alpha i}^a$. 
 It is emphasized that the auxiliary fermions occupy sites
 in the $D$-dimensional {\it spacetime} lattice,
 and they are different from the original fermions 
 that live on the $(D-1)$-dimensional space.
We define an inner product between quantum states defined in spacetime as
\beq
\langle  \Psi' || \Psi     \rangle\equiv    \langle \Psi'|O|\Psi\rangle
\eeq
with 
$O\equiv \prod_{i,a} ( c_{1i}^{a \dagger}- c_{1i}^a)( c_{2i}^{a \dagger} - c_{2i}^a )$.
The inner product is chosen such that the basis states satisfy the orthonormality condition,  
\beq
\la \psi', \bar \psi' || \bar\psi, \psi\ra = \prod_{i,a} \delta(\bar\psi_i^{'a}-\bar\psi_i^a)\delta(\psi_i^{'a}-\psi_i^a),
\label{inner_product2}
\eeq
where $\la \psi', \bar \psi' |$ is the Hermitian conjugate of $| \psi', \bar\psi'\ra$.
Then the partition function is given by the overlap between the two quantum states,
\beq
Z=\int D\bar\psi D\psi~ e^{-S_0[\bar\psi,\psi]-S_1[\bar\psi,\psi]} =\la S_0' || S_1\ra,
\eeq
where $|S_0'\ra$ is the complex conjugate of $|S_0\ra$.

Renormalization group flow can be understood 
as a quantum evolution of $|S_1\ra$
generated by an operator $\hat H$\cite{Lee2016}.
Since the partition function 
is invariant under the coarse-graining
transformation,
\beq
Z=\la S_0' || S_1\ra=\la S_0' || e^{-dz \hat H}| S_1\ra,
\eeq
where $dz$ is an infinitesimal parameter,
the reference state should be annihilated
by $\hat H$,
\beq
\hat H^\dagger O^\dagger |S_0'\ra=0.
\label{fixed}
\eeq
This is equivalent to the statement 
that $S_0$ represents a fixed point.
Because $|S_0'\ra$ is a direct-product state in spacetime,
the coarse graining operator 
that satisfies \eq{fixed}
is ultra-local,
\begin{equation}
\hat H= \sum_{i,a} 
\left[
-\frac 2 m c_{1i}^{a \dagger} c_{2i}^{a \dagger} 
+ c_{1i}^a c_{1i}^{a \dagger}
+ c_{2i}^a c_{2i}^{a \dagger}
\right].
\label{hamiltonian}
\end{equation}
The evolution generated by $\hat H$ corresponds to 
a real space coarse graining 
in which the mass $m$ is gradually increased 
such that fluctuations of fields are suppressed at each site\cite{Lunts2015}. 
While $\hat H$ is not Hermitian,
it can be mapped to a Hermitian operator through a similarity transformation 
as all eigenvalues are real.
As the evolution operators are repeatedly inserted between the overlap,
$e^{-z \hat H} |S_1 \ra$ is gradually projected to the 
unentangled ground state of $\hat H$.
Whether the initial state $|S_1\ra$ can be smoothly
projected to the direct product state 
depends on whether the system belongs to
the insulating phase described by the reference action\cite{Lee2016}.

Now we fix $\lambda$, 
and label the state associated with the deformation 
in terms of hopping amplitudes,
\beq
|t\rangle =\int D\bar\psi D\psi ~e^{\sum_{i,j}t_{ij}\bar\psi_i \cdot \psi_j-\frac \lambda N \sum_i(\bar\psi_i \cdot \psi_i)^2}|\bar\psi,\psi \rangle.
\eeq
Apart from the fixed quartic interaction,
$|t \rangle$ contains only single-trace hoppings.
However, multi-trace operators are generated
under the coarse graining evolution.
In quantum RG, the multi-trace operators are traded with 
non-trivial dynamics of the single-trace sources\cite{Lee2012,Lee:2012xba,Lee:2013dln}.
As a result, the partition function is given by 
a path integration over the scale dependent single-trace sources $t_{ij}(z)$\cite{Lunts2015,Lee2016},
\beqa
Z & =  & 
\lim_{z \rightarrow \infty}
 \la S_0^{'} || e^{- {z}  \hat H } | t^{(0)} \ra 
=  
\left.
\int Dt(z)  Dt^*(z)
 ~~ \la S_0^{'} || t(\infty)  \ra  ~ 
e^{ -N  \int_0^{\infty}  dz \left(
 t^*_{ij} \partial_z t_{ij} 
+     {\cal H}[t^*,t] 
\right) }
\right|_{t_{ij}(0)=t^{(0)}_{ij} }, \nn
\label{eq:Z}
\eeqa
where ${\cal H}[t^*,t] $ is the coherent state representation 
of the bulk Hamiltonian $\hat {\mathcal H}$ given by
\beqa
\hat{\mathcal H}& =& \sum_i  \Big[  \frac{2}{m}t_{ii}+\frac{4\lambda(-1+\frac 1 N)}{m}t_{ii}^\dagger-4\lambda(t_{ii}^\dagger)^2-\frac{8\lambda^2}{m}(t_{ii}^\dagger)^3     \Big]    \nn
&&   +\sum_{i,j}\Big[ 2+\frac {4\lambda}{m}(t_{ii}^\dagger+t_{jj}^\dagger)    \Big] t_{ij}^\dagger t_{ij}-  \frac{2}{m}\sum_{i,j,k} \Big[ t_{kj}^\dagger t_{ki} t_{ij}   \Big].
\eeqa
Here the gauge is fixed so that 
the speed of coarse graining is uniform in spacetime,
and the shift is zero at all $z$\cite{Lee2016}.
The bi-local fields are the fundamental degrees of freedom in the bulk,
which include the metric and higher spin fields.
In the Hamiltonian picture, 
the bi-local fields are promoted to quantum operators :
$t_{ij}$ ($t_{ij}^\dagger$) represents 
the annihilation (creation) operator
of quantized link between sites $i$ and $j$.
Despite the fact that the original theory is fermionic,
the bulk theory is bosonic 
because there is no U(N) invariant 
fermionic operator in the theory.

In the large $N$ limit, 
quantum fluctuations in the RG path 
are suppressed,
and the saddle point approximation becomes reliable.
The saddle point equation reads
\begin{eqnarray}
\begin{aligned}
\partial_z \bar t_{ij}&=-2\Big\{ -\frac{2\lambda \delta_{ij}}{m}-\delta_{ij}[4\lambda+\frac{12\lambda^2}{m}\bar p_{ii}]\bar p_{ii}  +\frac{2\lambda \delta_{ij}}{m}\sum_k (\bar t_{ik}\bar p_{ik}+ \bar t_{ki} \bar p_{ki})\\
&+[1+\frac{2\lambda}{m}(\bar p_{ii}+\bar p_{jj})]\bar t_{ij} -\frac{1}{m}\sum_k \bar t_{ik} \bar t_{kj} \Big\},\\
\partial_z \bar p_{ij}&=2\Big\{ \frac{\delta_{ij}}{m}+[1+\frac{2\lambda}{m}(\bar p_{ii}+\bar p_{jj})]\bar p_{ij}  -\frac{1}{m}\sum_k (\bar p_{ik}\bar t_{jk}+ \bar t_{ki} \bar p_{kj}) \Big\}.
\label{eomreal}
\end{aligned}
\end{eqnarray}
Here $\bar t_{ij}(z)$ and $\bar p_{ij}(z)$ denote the saddle point configuration
of $t_{ij}(z)$ and $t_{ij}^*(z)$, which satisfy the boundary conditions,
\begin{equation}
\bar t_{ij}(0)=t_{ij}^{(0)},
\end{equation}
\begin{equation}
\bar p_{ij} (\infty)=\frac{1}{N} \left. \frac{\partial ln \langle S_0^{'} || \bar t \rangle}{\partial \bar t_{ij}} \right|_{z=\infty}.
\end{equation}
It is noted that $\bar p_{ij}$ is not necessarily the complex conjugate of $\bar t_{ij}$
at the saddle point.
If $t_{ij}^{(0)}$ depends only on $r_i - r_j$,
the saddle point solution is also invariant under the translation.
In this case, the equations in momentum space are reduced to 
\begin{eqnarray}
\partial_z   \bar T_{q} &=&
-2 \Bigg\{
-\frac{2\lambda}{m} 
-  \left[ 4\lambda + \frac{12\lambda^2}{m}  \bar p_{0} \right]  \bar p_{0} 
+ \frac{4\lambda}{V m} \sum_{q^{'}}  \bar T_{q^{'}}  \bar P_{q^{'}}  + 
\left[ 1+\frac{4\lambda}{m}   \bar p_{0} \right]  \bar T_{q}
-\frac{1}{m}   (\bar T_{q})^2 \Bigg\}, \label{EOMP1} \\
\partial_z  \bar P_{q} &=&
\frac{2}{m}
+2 \left[1+\frac{4\lambda}{m}  \bar p_{0} \right]  \bar P_{q}
-\frac{4}{m}  \bar P_{q}   \bar T_{q},
\label{EOMP2} 
\end{eqnarray}
where $\bar T_q=\sum_r \bar t_{i+r,i}e^{iqr}$, 
$\bar P_q=\sum_r \bar p_{i+r,i}e^{-iqr}$, $\bar p_0\equiv \bar p_{ii}=\frac 1 V \sum_q \bar P_q$.
$q = (\omega, k)$  denotes frequency, $\omega$ 
and $(D-1)$-dimensional momentum, $k=(k_1,k_2,..,k_{D-1})$.
The solution to the saddle point equations is given by 
\begin{equation}
\bar P_q(z)=-\frac {e^{-2z}}{- i\bar T_q(0)+m+2\lambda p_0(0)}-(1-e^{-2z})\frac{1}{m},
\label{21}
\end{equation}
\begin{equation}
\bar T_q(z)=-\frac {2 \lambda}{m}+m+\frac {2\lambda}{m}e^{-2z}(m \bar p_0(0)+1)+\frac {1}{\bar P_q(z)}.
\label{22}
\end{equation}

Eqs. (\ref{21}) and (\ref{22}) completely determine
the saddle-point configurations of the bulk fields 
in terms of $\bar T_q(0)$,
which encodes all the information about the microscopic theory.
For example, the $(D-1)$-dimensional hypercubic lattice with a continuous imaginary time gives 
$\bar T_q(0) = - i \omega - 2 t \sum_{i=1}^{D-1} \cos ( k_i ) - \mu$,
where $t$ is the nearest neighbor hopping amplitude
and $\mu$ is the chemical potential.
In general, any local hopping in space 
can be written in power series of $k$,
\beq
\bar T_q(0)=\bar T_0(0)-i\omega - e(k),
\label{eq:dispersion}
\eeq
where 
\beq
e(k) = \bm k^2 + \sum_{n=2}^\infty 
\frac{c_{i_1,i_2,..,i_{2n}} }{ m^{n-1} }
k_{i_1} k_{i_2}..k_{i_{2n}}.
\label{eq:dispersion2}
\eeq
Here the chemical potential $\bar T_0(0)$ is singled out,
and the coefficient of the quadratic term is normalized by rescaling $k$.
$ c_{i_1,i_2,..,i_{2n}} $ can be independently tuned by further neighbor hoppings
on a microscopic lattice. 
It is assumed that 
the minimum of the band is $k=0$,
and 
the lattice 
respects the parity
to allow only even powers of momentum in the dispersion.
In terms of $e(k)$, the saddle point solutions are written as
\beqa
\bar P_q(z) & =& -\frac {e^{-2z}}{i\omega+ e(k) +\delta}-(1-e^{-2z})\frac{1}{m}, \nn
\bar T_q(z) & =& -\frac {2 \lambda}{m}+m+\frac {2\lambda}{m}e^{-2z}(m \bar p_0(0)+1)-m \frac {i\omega+ e(k) +\delta}{(1-e^{-2z})(i\omega+ e(k) +\delta)+me^{-2z}},
\label{Tqz}
\eeqa
where $\de =-\bar T_0(0)+m+2\lambda\bar p_0(0)$ is the gap.


\section{Emergent geometry in the bulk}

Because the reference state is ultra-local,
there is no background metric in the bulk.
The bulk geometry is dynamically determined by 
the saddle point solution.
Since the bulk theory involves the bi-local fields of all sizes,
it is kinematically non-local.
A sense of locality emerges only when
single-trace sources decay
fast enough in spacetime. 
In this section, we examine the geometry 
that emerges in the bulk in the insulating phase and in the metallic phase.

\subsection{Insulating phase}

To study the behavior of the hopping field in real space, 
we first transform $\bar T_q(z)$ to the time domain,
\begin{equation}
\bar T_{\tau,\bm k}(z)=\int \bar T_{\omega,k}(z)e^{-i\om \tau}\frac{d\om}{2\pi}= \frac{m^2e^{2z}}{(e^{2z}-1)^2} 
\Bigl[ -\theta(\tau)\theta(-E(\bm k,z))+\theta(-\tau)\theta(E(\bm k,z)) \Bigr] 
e^{E(\bm k,z) \tau},
\label{Ttaukz}
\end{equation}
where 
\begin{equation}
E(\bm k,z)= e(k) + \delta+\f{m}{e^{2z}-1}.
\end{equation}
$\theta(\tau)$ is the theta function,
and ultra-local terms are ignored. 
In the insulating phase with $\de>0$, 
$E(\bm k,z)$ is positive
for all $k$ and $z$,
and the first term on the right hand side of \eq{Ttaukz} vanishes.
Because $E(\bm k,z)$  is analytic in $k$,
the hopping field decays exponentially in real space at all $z$.
For example, for $e(k)=k^2$ the bi-local field in the real space is given by
\begin{equation}
\bar t_{i+r,i}(z)= \int \bar T_{\tau,\bm k}(z)e^{-i\bm k\cdot\bm x}\frac{d\bm k}{(2\pi)^{d-1}}=
\left(\f{1}{2\sqrt{-\pi\tau}}\right)^{d-1}
\frac{m^2e^{2z}}{(e^{2z}-1)^2}
\theta(-\tau)e^{(\delta+\f{m}{e^{2z}-1})\tau}e^{\f{\bm x^2}{4\tau}},
\label{Ttauxz_local}
\end{equation}
where $r=(\tau,x)$
with  $x=(x_1,x_2,..,x_{D-1})$ 
representing $(D-1)$-dimensional vector in real space.


Fluctuations of the bi-local fields propagate  
on the background set by the saddle-point configuration.
Therefore, the bulk geometry can be extracted 
by inspecting the equation of motions 
obeyed by the fluctuations of the hopping fields around the saddle point,
\beq
\blgn
\td t_{ij}&=t_{ij}-\bar t_{ij},\\
\td p_{ij}&=t_{ij}^*-\bar p_{ij}.
\elgn
\eeq
The quadratic action for the fluctuations is
\beq
\blgn
S_2 & =  \int dz \Bigg\{
-4 \lambda \sum_i 
\left( 1+ \frac{6 \lambda}{m} \bar p_0 \right) \td p_{ii}^2 \\
& +  \sum_{ij} \left[ 
\td p_{ij} \left( \partial_z  + 2 + \frac{8 \lambda}{m} \bar p_0 \right) \td t_{ij} 
 + \frac{4 \lambda}{m}
(\td p_{ii} + \td p_{jj} ) 
\left( 
\bar p_{ij} \td t_{ij}
+ \bar t_{ij} \td p_{ij}
\right) \right] \\
& - \frac{2}{m} \sum_{ijk} 
\left(
\bar p_{ij} \td t_{ik} \td t_{kj}
+  \bar t_{ik}  \td p_{ij} \td t_{kj}
+  \bar t_{kj}  \td p_{ij} \td t_{ik}
\right)
\Bigg\}.
\label{eq:bulkaction}
\elgn
\eeq
To extract the background metric, 
we focus on the imaginary part of the hopping field 
$\td t_{ij}^A= \td t_{ij}-\td t_{ij}^*$ 
which satisfies a simple equation of motion,
\beq
\left( \partial_z + 2 + \frac{ 8 \lambda}{m} \bar p_0(z) \right) \td t^A_{ij}
 - \frac{2}{m} \left( 
\bar t_{ik} \td t^A_{kj}  + \bar t_{kj} \td t^A_{ik} \right)  = 0
\label{QEOMA} 
\eeq
for $i \neq j$.
We take the continuum limit, 
and define $\td t_{ij}^A=e^{-2z}\td t^A(r,r',z)$, 
where $r$ and $r'$ represent spacetime coordinates $(\tau, \bm x)$ and $(\tau ', \bm x')$.
For large $z$
and $r \neq r'$, the equation of motion becomes
\beq
\left[ 
 ( m+\de e^{2z})  \pa_z + 4 \delta  
-   \f{2m}{1-\frac{\pa_\tau+ e( \partial_i ) }{me^{-2z}+\de}}  
-\f{2m}{1-\frac{ -\pa_\tau '+ e( \partial_i^{'} )  }{me^{-2z}+\de}}          
 \right]  \td t^A(r,r',z)=0,
\label{EOMA}
\eeq
where 
$\partial_i \equiv \frac{\partial}{\partial x_i}$,
$\partial_i^{'} \equiv \frac{\partial}{\partial x_i^{'}}$.
This can be written in a covariant form,
\beqa
\Bigg[
    \sqrt{g^{zz}(z)} \pa_z + 4 \delta 
    -\f{2m}{1- \frac{1}{m} \left(\sqrt{g^{\tau\tau}(z)}\pa_\tau + g^{ij}(z)\pa_i \pa_j 
+  \sum_{n>1} \frac{\td c_{i_1,..,i_{2n}}(z)}{m^{n-1}} \pa_{i_1}..\pa_{i_{2n}} \right)} \nn
    -\f{2m}{1- \frac{1}{m} \left(- \sqrt{g^{\tau\tau}(z)}\pa_{\tau'} +g^{ij}(z)\pa_i' \pa_j' 
+  \sum_{n>1} \frac{\td c_{i_1,..,i_{2n}}(z)}{m^{n-1}} \pa_{i_1}'..\pa_{i_{2n}}' \right)} 
  \Bigg] \td t^A(r,r',z)=0,
\label{FL}
\eeqa
where $\sqrt{g^{zz}}(z)=m+\de e^{2z}$, 
$\sqrt{g^{\tau\tau}}(z)=g^{ii}(z)=\f{m}{m e^{-2z}+\de}$,
$\td c_{i_1,..,i_{2n}}(z) = \frac{ m  }{m e^{-2z}+\de} c_{i_1,..,i_{2n}}$. 
The bulk metric is identified from the terms up to two derivatives,
\beq
ds^2=\f{dz^2}{(m+\de e^{2z})^2}+(e^{-2z}+\de/m)^2 d\tau^2+(e^{-2z}+\de/m)d\bm x^2.
\label{eq:metric}
\eeq
$\td c_{i_1,..,i_{2n}}(z)$ encodes 
the vacuum expectation values of the
higher spin fields in the bulk.

In the insulating phase,
the geometry ends at a finite proper distance from the UV boundary, 
\beq
L=\int_0^\infty \f{dz}{m+\de e^{2z}}=\f{log(\f{m+\de}{\de})}{2m}.
\eeq
The finite depth in the bulk reflects the fact that 
the quantum state associated with the action 
can be smoothly projected to the direct product state
through a series of RG transformations with a finite depth. 
The proper distance in the radial direction
measures the ``distance'' between theories\cite{Lee2016}.

\subsection{Metallic phase}

As the system deviates further from the insulating state with decreasing gap,
the depth of the bulk space diverges logarithmically.
When the system reaches the critical point at $\de=0$,
the bulk geometry develops an infinitely long throat
with a Lifhistz horizon at $z=\infty$.
At the critical point, the metric in \eq{eq:metric}
becomes the Lifshitz geometry,
\beq
ds^2=\f {dz^2}{m^2}+e^{-4z}d\tau^2+e^{-2z}d\bm x^2
\label{eq:Lifshitz}
\eeq
which is invariant under the scaling
\beq
z\to z+s,~\tau\to e^{2s} \tau,~\bm x\to e^s \bm x.
\label{eq:ss}
\eeq
The dynamical critical exponent $z=2$ 
reflects the quadratic dispersion at the bottom of the band 
in \eq{eq:dispersion}.
The isometry of the bulk metric is expected from the form of
the hopping field in \eq{Ttauxz_local}
which is invariant under \eq{eq:ss}
at $\de=0$.
If one tuned parameters in $t_{ij}$ at UV
to make the dispersion to scale as $\bar T_q(0) \sim k^{2r}$
at the bottom of the band,
the resulting geometry would have the dynamical critical exponent $2r$.
It is noted that the higher-derivative terms in \eq{FL}
are suppressed by the curvature scale in the bulk.
For the locality at a shorter length scale in the bulk,
there are stringent constraints 
on field theories\cite{Heemskerk:2009pn,Nakayama2015,2016arXiv161105315S}.

At the critical point,
the Lifshitz horizon is located at $z=\infty$.
However, the horizon arises at a finite $z$ in the metallic phase.  
Unlike the bosonic model\cite{Lee2016}, 
the chemical potential can be further increased 
across the bottom of the band.
For $\de < 0$, a Fermi surface forms,
 and the system becomes a metal.
In the metallic phase, 
the horizon moves to $z_H\equiv\f 1 2 log(1+\f{m}{-\de})$
at which $g_{\tau \tau}$ vanishes.
Outside the horizon, $E(\bm k,z)$ remains positive for all $k$,
and the metric is given by
\beq
ds^2=\f{dz^2}{4m^2 (z_H-z)^2}+\f{4\de^2}{m^2}(z_H-z)^2 d\tau^2+\frac{-2\de}{m}(z_H-z)d \bm x^2
\label{eq:outH}
\eeq
for small $z_H-z$ with $z_H >> 1$.
The geometry no longer has an isometry 
associated with a global translation of $z$
because there is a scale represented by $z_H$.
However, the space outside the horizon 
has an isometry associated with rescaling of $z$ toward $z_H$.
The isometry becomes manifest in a new coordinate system, 
$z'=-\f 1 2 log(z_H-z)$, 
$\tau '= -\f{2\de}{m} \tau$, 
$\bm x'=\sqrt{\f{-2\de}{m}}\bm x$
in which the metric reduces to the Lifshitz geometry
in \eq{eq:Lifshitz}.
Although the horizon arises at a finite $z$, 
the proper distance from the UV boundary to the horizon is still infinite.
This is consistent with the fact that 
the metallic state belongs to a different universality class from the insulating state.

Although the metric near the horizon takes the universal form,
the background higher-spin fields, $\td c_{i_1,..,i_{2n}}(z)$ 
encodes the information about the microscopic details 
such as the underlying lattice and further neighbor hoppings 
in the boundary theory\cite{PROP:PROP200410203,Grumiller2016,PhysRevLett.116.231301}. 
The effect of the higher-spin hair on the fluctuation fields in \eq{FL} 
depends on the momentum of the fluctuation mode and the radial position.
For a mode whose proper momentum is  $O(\sqrt{m})$ 
at radial location $z$,
the $2n$ derivative term in \eq{FL} scales as $(z_H - z)^{(n-1)}$.
The bulk equation of motion for the low-momentum modes
becomes insensitive to the higher-spin fields close to the horizon.
However, in order to extract the full low-energy data from the UV boundary,
one should probe particle-hole excitations 
whose momenta are comparable 
to the Fermi momentum $k_F \sim \sqrt{|\delta|}$.
If a mode with proper momentum $k_F$ at the UV boundary
is sent into the bulk, 
the momentum is blue shifted to 
$p \sim  \sqrt{\frac{|\delta|}{ (z_H-z)}}$ near the horizon.
For these modes,
the higher-derivative terms in \eq{FL} 
remain important.
The sensitivity to the higher-spin fields 
signifies the need to go beyond the low-energy effective theory near the horizon.
This is expected because
the shape of Fermi surface is determined 
not just by the quadratic term but also 
by all higher-order terms in \eq{eq:dispersion2}.

The horizon corresponds to a phase transition
at which the length scale associated 
with the size of bi-local operators in $e^{-z \hat H} | S_1 \ra$ 
diverges\cite{Lee2016}.
At the horizon, the range of hoping diverges, 
which causes the collapse of the $D$-dimensional spacetime volume.
Zero area of the horizon is consistent with the fact that
the metal has zero entropy density.
Although the Lifshitz horizon has a divergent tidal force,
nothing stops one from being able to continue 
the coarse graining procedure across the horizon. 
Inside the horizon with $z>z_H$,
$E(\bm k,z)$ changes sign as a function of momentum.
The discontinuity of 
$\bar T_{\tau,\bm k}(z)$ as a function of momentum 
results in a slow decay of the hopping field 
along with a Friedel-like oscillation in real space. 
The specific form of the oscillation depends on the microscopic details.
From now on, we will focus on the dispersion $e(k) = k^2$,
which describes the spherical Fermi surface.
For different shapes of Fermi surface, 
the structure inside the horizon will be different.
For the spherical Fermi surface, 
$E(\bm k,z)$ becomes negative for
$|k|< k_s(z)  \equiv\sqrt{-\de-\f{m}{e^{2z}-1}}$,
\begin{equation}
\bar T_{\tau,\bm k}(z)= 
\frac{m^2e^{2z}}{(e^{2z}-1)^2}
\Big(-\theta(\tau)\theta(k_s(z)-|\bm k|)+\theta(-\tau)\theta(|\bm k|-k_s(z))\Big) e^{(\bm k^2-k_s(z)^2)\tau}.
\label{Ttaukz_metallic}
\end{equation}
The asymptotic behavior of $\bar t_{i+r,i}$ 
as a function of $r=(\tau,x)$ is given by
$t_{i+r,i} = \frac{m^2 e^{2z} }{(e^{2z}-1)^2} K(\tau,x;z)$,
where
\begin{equation}
K(\tau,x;z) \sim -  \f{\cos(k_s(z) x)}{2\pi k_s(z) \tau}+O(\f{1}{\tau^2}), \textrm{   as }|\tau|\to \infty,
\label{25}
\end{equation}
\begin{equation}
K(\tau,x;z) \sim -  \f{\sin(k_s(z) x)}{\pi x}+O(\f{1}{x^2}), \textrm{   as }x\to \infty
\label{26}
\end{equation}
in $D=1+1$, and 
\begin{equation}
K(\tau,x;z) \sim -   \frac{1}{8\pi^2 \tau}\cdot J_0(k_s(z) |x|)+O(\f{1}{\tau^2}), \textrm{   as }|\tau|\to \infty,
\label{27}
\end{equation}
\begin{equation}
K(\tau,x;z) \sim -   \sqrt{\f {k_s(z)}{2 \pi^3 |\bm x|^3}}\cdot \sin(k_s(z) |\bm x| -\f \pi 4)+O(|\bm x|^{-\f 5 2}), \textrm{   as }\bm x\to \infty
\label{28}
\end{equation}
in $D=2+1$, where $J_0(x)$ is the 0th order Bessel function of the first kind.

Inside the horizon, the saddle point configuration 
of the hopping field decays in a power-law as a function of $r$.
The hopping field  sets a non-local background on which 
fluctuation fields propagate according to \eq{QEOMA}. 
In the large $z$ limit, 
the equation of motion for 
$\td t^A(\tau,x,\tau',x',z)$ becomes
\beqa
  && ( m e^{2z} ~ \pa_z + 4m )  \td t^A(\tau,x,\tau',x',z)    -
2 m^2 \int d \tau_1 dx_1 ~ K(\tau-\tau_1,x-x_1;z) ~\td t^A(\tau_1,x_1,\tau',x',z)  \nn
&& ~~~~~~ - 2 m^2 \int d\tau'_1 dx'_1 
~ \td t^A(\tau,x,\tau'_1,x'_1,z)    
~K(\tau'_1-\tau',x'_1-x';z) 
=0,
\label{eq:inside}
\eeqa
where the asymptotic behavior of $K(\tau,x;z)$ is given in 
Eqs.(\ref{25})-(\ref{28}).
The non-local saddle point configuration 
allows the bi-local fields to jump to far sites
in the $D$-dimensional spacetime.
This implies that the space inside the horizon can not 
be described by a local geometry.

The wavevector for the Friedel-like oscillation at radial position $z$
is given by $k_s(z)$,
which gradually increases from $0$ at $z_H$
to the true Fermi momentum $k_F = \sqrt{-\de}$ in the large $z$ limit.
$k_s(z)$ can be regarded as the size of the Fermi surface
with a $z$-dependent chemical potential 
$\mu(z) = -\de-\f{m}{e^{2z}-1}$.
As $z$ increases from the horizon,
the bulk effectively scans through the occupied states
from the bottom of the band to the Fermi level.
Therefore, the non-local structure inside the horizon
is determined by the full dispersion below the Fermi surface.
The region deep inside the horizon
keeps the data on the low-energy modes close to the Fermi surface.
Different metals have different non-local structures
in the large $z$ limit because the Fermi liquids have 
infinitely many marginal deformations, including the shape of the Fermi surface.
On the other hand, the region close to the horizon in the interior
is sensitive to the occupied states below the Fermi surface,
which is not part of the universal low-energy data
of the boundary theory.
A change in the shape of the band below the Fermi surface 
that does not involve a change near the Fermi surface
does affect the wavevector of the oscillation near the horizon. 
This implies that  
the interior of the horizon
keeps not only the universal low-energy data of the theory
but also  non-universal microscopic information 
of the boundary field theory.

\section{Summary and Discussion}

In this paper, we apply the quantum renormalization group
to construct the holographic dual of the fermionic $U(N)$ vector model
with a nonzero charge density.
The bulk equations of motion are exactly solved in the large $N$ limit 
to derive the geometries that emerge in different phases of the model.
In the insulating phase, the bulk geometry caps off at a finite proper distance
in the radial direction.
At the critical point between the metal to insulator phase transition,
the proper size in the radial direction diverges,
which results in a Lifshitz geometry in the bulk.
In the metallic phase, the Lifshitz horizon
moves to a finite radial coordinate. 
The geometry outside of the Lifshitz horizon carries 
a higher-spin hair determined 
by microscopic hopping amplitudes 
of the boundary theory.
On the other hand, the interior of the horizon is 
characterized by an algebraic non-locality,
and can not be described by a Riemannian geometry.
Certain microscopic data of the boundary field theory
is encoded in the interior of the horizon.

\begin{figure}[h]
\centering
\includegraphics[width = 0.5\columnwidth]{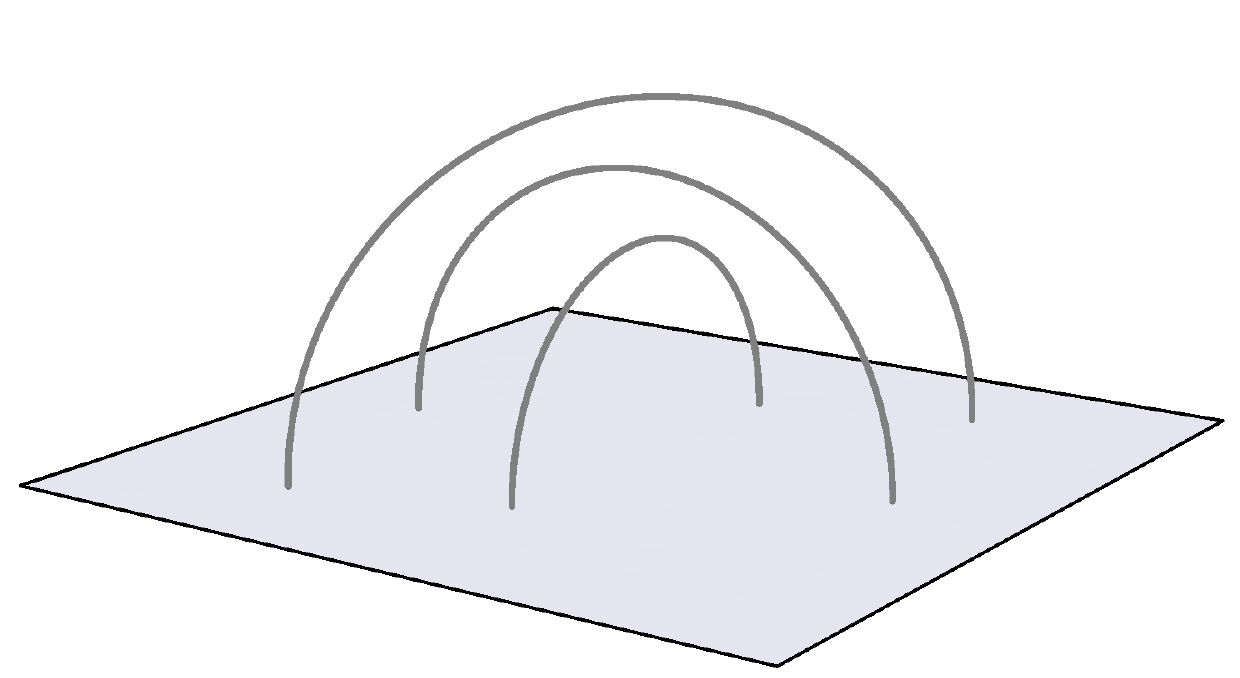}
\caption{
Examples of geodesics in the 
three-dimensional hyperbolic space ${\bf H}_3$.
The shortest geodesic distances between two points
on the plane scales logarithmically 
in the coordinate distance in the plane.
}
\label{hyperbolic}
\end{figure}

The non-local space inside the horizon 
is different from a globally connected network
where every site is connected to every other sites 
with an equal hopping strength.
Since the non-local connectivity decreases as a power-law
in  coordinate distance,
there is still a sense in which 
certain points are `closer' than other points to a given point.
However, this notion of distance can not be captured 
within the framework of  Riemannian geometry.
In order to define a notion of distance in this space,
one might rewrite the non-local kernel in \eq{eq:inside} as
\beqa
  K(\tau,x) =   \tilde K(\tau,x) e^{ - m d(\tau-\tau_1,x-x_1)  },
 \label{eq:inside2}
\eeqa
where $\tilde K( \tau,x)$ is a function which captures
the modulation in the sign of the hopping function with $|\tilde K(\tau,x)| \sim 1$
and  $d(\tau,x)$ is a function that measures physical distance.
The idea behind this is to define the physical distance  
by insisting that the kernel in the kinetic term decays exponentially.
This agrees with the geodesic distance measured by the metric in \eq{eq:outH}
outside the horizon, 
but the definition is general enough to be applicable to the region inside the horizon.
Because the hopping amplitude decays in power-law in \eq{26} and \eq{28},
the distance between two points increases only logarithmically in coordinate,
e.g. $d(x,y) \sim \frac{1}{m} \log |x-y|$.
No metric on a Riemannian manifold 
can reproduce the distance function of this form.
One way of obtaining such distance function is 
by embedding the space in a higher dimensional
Riemannian space, and define the distance 
between two points on the original space 
in terms of geodesics that can go through the higher dimensional space.
For example, the logarithmic distance function can be obtained
from the geodesic distance between points 
on the boundary of a hyperbolic space (see Fig. \ref{hyperbolic}). 
The non-local geometry is `anomalous' in that
distance function can be realized from a local metric 
only through a higher-dimensional space.

For $D > 2$, 
the metallic state becomes unstable
against superconductivity 
at an energy scale which is exponentially small 
in $N$.
As a result, the geometry for the metallic state
becomes unstable once quantum fluctuations
are included in the bulk.
However, this should not be the case for all metals.
If one breaks the parity and the time-reversal symmetry,
one can have a Fermi liquid state 
without perturbative superconducting instability.
The geometries dual to those stable metals 
are expected to exhibit similar 
non-local structures behind the horizon.

Since we use the imaginary time formalism in this study,
the bulk theory captures only the ground state of the theory.
It shows how bulk geometry depends on 
the microscopic information of the ground state
determined by Hamiltonian. 
It will be interesting 
to extend this to the real time formalism
to understand the dependence of bulk geometry
on state and Hamiltonian separately.

\section*{Acknowledgments}

SL thanks the participants of the 
Simons symposiums on quantum entanglement
and the Aspen workshop on emergent spacetime in string theory
for inspiring discussions.
The research was supported by 
the Natural Sciences and Engineering Research Council of Canada.
Research at the Perimeter Institute is supported 
in part by the Government of Canada 
through Industry Canada, 
and by the Province of Ontario through the
Ministry of Research and Information.

\newpage
\appendix


\section{Asymptotic Behavior of $\bar T_{\tau,x}$}
\label{asymptotic}

In this appendix, we examine the asymptotic behavior of 
$\bar t_{i+r,i}$ as $|\tau|\to\infty$ and $x\to\infty$ in $D=1+1$ and $D=2+1$.
In $1+1$ dimensions, the Fourier transform of \eq{Ttaukz_metallic} gives 
\begin{equation}
\begin{aligned}
\bar t_{i+r,i}(z)=\frac{m^2e^{2z}}{(e^{2z}-1)^2}\cdot e^{-k_s^2\tau}\cdot   \f{e^{\f{x^2}{4\tau}}}{4\sqrt{\pi}} \Big[ - \theta(\tau)\f{i}{\sqrt{\tau}} \Big(erf(\f{x-2ik_s \tau}{2\sqrt{\tau}})-erf(\f{x+2ik_s \tau}{2\sqrt{\tau}})\Big)  \\
+\theta(-\tau) \f{1}{\sqrt{-\tau}} \Big(2-erf(\f{ix-2k_s \tau}{2\sqrt{-\tau}})+erf(\f{ix+2k_s \tau}{2\sqrt{-\tau}})\Big)   \Big],
\label{Ttauxz_metallic}
\end{aligned}
\end{equation}
where $erf(x)$ is the error function. 
The asymptotic form is given by \eq{25} and \eq{26}.


\begin{figure}[h]
\centering
\includegraphics[width = 0.5\columnwidth]{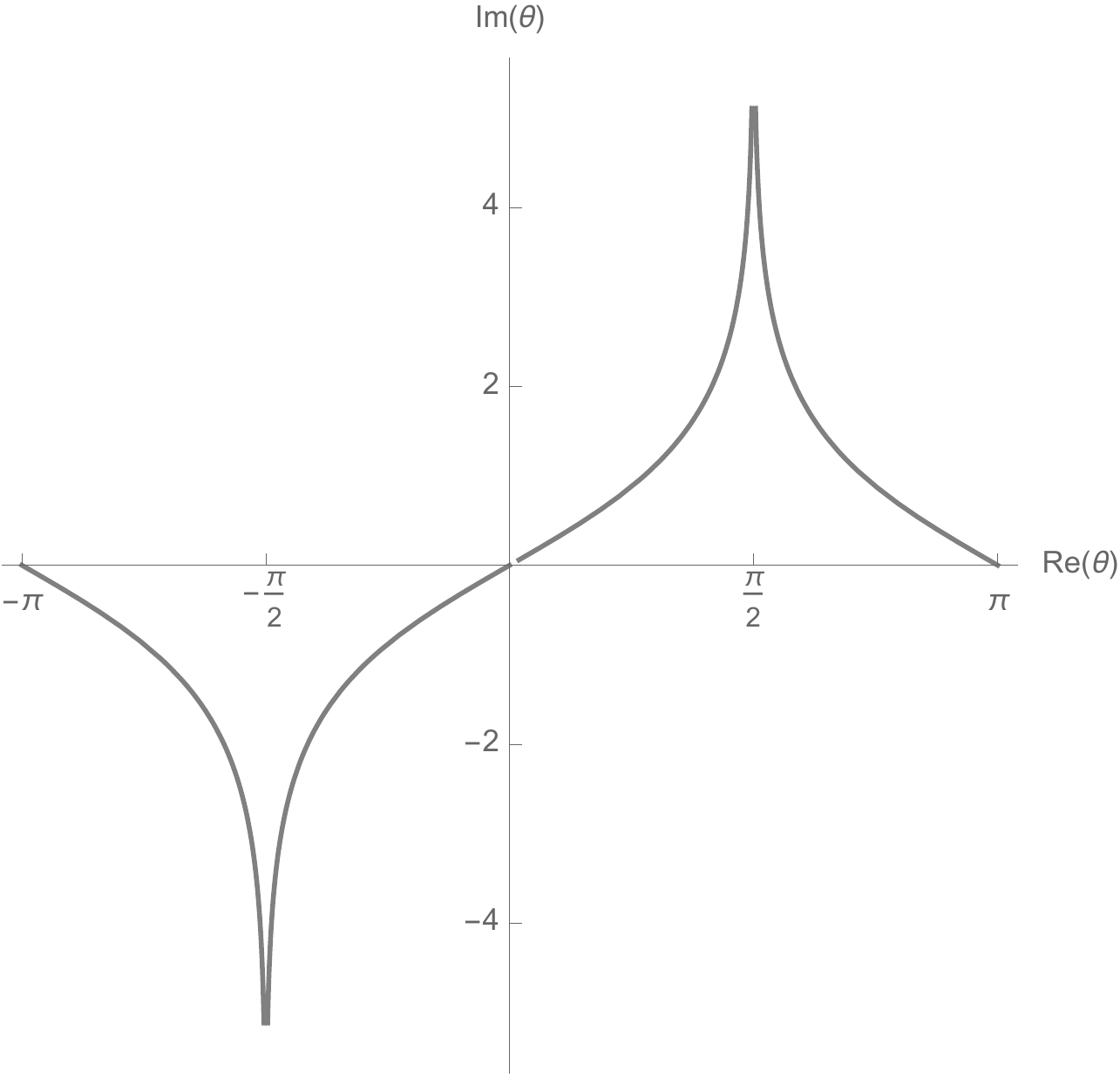}
\caption{integral contour of $\theta$ where Re$(\cos\theta)=\pm 1$}
\label{steepest_descent}
\end{figure}

In $2+1$ dimensions, the Fourier transformation of \eq{Ttaukz_metallic} becomes
\begin{equation}
\begin{aligned}
\bar t_{i+r,i}(z)=& \int_{-\pi}^{\pi}\int_0^\infty \bar T_{\tau,\bm k}(z)e^{-ikxcos\theta}\cdot k\cdot \frac{d kd\theta}{(2\pi)^{2}}.
\end{aligned}
\end{equation}
For large $|\tau|$, \eq{Ttaukz_metallic} shows that 
$\bar T_{\tau,k}(z)$ falls very fast as $|k|$ goes away from $k_s$. 
Since the integration over $k$ is concentrated near $k_s$,
we obtain
\begin{equation}
\begin{aligned}
\int_0^\infty \bar T_{\tau,\bm k}(z)e^{-ikxcos\theta}\cdot k\cdot \frac{d k}{2\pi}
&=-\frac{m^2 e^{2z}}{(e^{2z}-1)^2}\int_{0}^{k_s}e^{(k^2-k_s^2)\tau}e^{-ikxcos\theta}\cdot k\cdot \frac{dk}{2\pi}\\
&\approx-\frac{m^2 e^{2z}}{(e^{2z}-1)^2} \int_{-\infty}^{k_s}k_s\cdot e^{2k_s(k-k_s)\tau}e^{-ik_sxcos\theta}\frac{dk}{2\pi}    \\
&=-\frac{m^2 e^{2z}}{(e^{2z}-1)^2}\cdot \frac{1}{4\pi \tau}\cdot e^{-ik_sxcos\theta},
\label{positive_tau}
\end{aligned}
\end{equation}
and 
\begin{equation}
\begin{aligned}
\bar t_{i+r,i}\approx &-\frac{m^2 e^{2z}}{(e^{2z}-1)^2}\cdot \frac{1}{8\pi^2 \tau}\int_{-\pi}^{\pi} e^{-ik_sxcos\theta}d\theta\\
=&-\frac{m^2 e^{2z}}{(e^{2z}-1)^2}\cdot \frac{1}{8\pi^2 \tau}\cdot J_0(k_sx).
\end{aligned}
\end{equation}
On the other hand, for large $|x|$, 
we first perform the radial integration in momentum to obtain
\begin{equation}
\begin{aligned}
\int_0^\infty \bar T_{\tau,\bm k}(z)e^{-ikxcos\theta}\cdot k\cdot \frac{d k}{2\pi}\approx \frac{m^2 e^{2z}}{(e^{2z}-1)^2}\cdot \frac{k_s e^{ -ik_s xcos\theta}}{ 2 i\pi xcos\theta}.
\end{aligned}
\end{equation}
To evaluate the remaining angle integration,  
\begin{equation}
\begin{aligned}
\bar t_{i+r,i}\approx &\frac{m^2 e^{2z}}{(e^{2z}-1)^2}\cdot \frac{k_s}{4i\pi^2 }\int_{-\pi}^{\pi} \frac{ e^{ -ik_s xcos\theta}}{xcos\theta}d\theta
\end{aligned}
\end{equation}
we use the method of steepest descent for large $|x|$,
where the integral contour is deformed as in Fig. \ref{steepest_descent}.
The main contributions are from $\theta=\pm \pi$ and $\theta=0$. 
Expansions around these points give \eq{28}.

\bibliography{references}
\bibliographystyle{JHEP}

\end{document}